\documentclass[twocolumn,aps,prl,superscriptaddress,amsmath,amssymb,floats]{revtex4}
\usepackage{graphicx}
\usepackage{amsmath}
\usepackage{amssymb}
\usepackage{graphicx}
\usepackage{CJK}
\usepackage{keyval}
\usepackage{dcolumn}
\usepackage{bm}
\usepackage{color}
\usepackage{txfonts}
\usepackage[]{sidecap}

\begin{document}

\title{Distinct Kondo Screening Behaviors in Heavy Fermion Filled Skutterudites with $4f^1$ and $4f^2$ Configurations}

\author{X. Lou}

\author{H. C. Xu}
\author{T. L. Yu}
\author{Y. H. Song}
\author{C. H. P. Wen}
\author{W. Z. Wei}

\affiliation{%
Laboratory of Advanced Materials, State Key Laboratory of Surface Physics, and Department of Physics, Fudan University, Shanghai 200438, China
}%


\author{A. Leithe-Jasper}
\affiliation{%
Max-Planck-Institut f$\ddot{u}$r Chemische Physik fester Stoffe, N$\ddot{o}$thnitzer Stra$\beta$e 40, 01187 Dresden, Germany
}%
\author{Z. F. Ding}
\affiliation{%
Laboratory of Advanced Materials, State Key Laboratory of Surface Physics, and Department of Physics, Fudan University, Shanghai 200438, China
}%
\author{L. Shu}
\affiliation{%
Laboratory of Advanced Materials, State Key Laboratory of Surface Physics, and Department of Physics, Fudan University, Shanghai 200438, China
}%
\affiliation{%
Shanghai Research Center for Quantum Sciences, Shanghai 201315, China
}%

\author{S. Kirchner}

\affiliation{%
Zhejiang Institute of Modern Physics and Department of Physics, Zhejiang University, Hangzhou, 310027, China
}%
\affiliation{%
Zhejiang Province Key Laboratory of Quantum Technology and Device, Zhejiang University, Hangzhou 310027, China
}%

\author{R. Peng}
\email{pengrui@fudan.edu.cn}
\affiliation{%
Laboratory of Advanced Materials, State Key Laboratory of Surface Physics, and Department of Physics, Fudan University, Shanghai 200438, China
}%
\affiliation{%
Shanghai Research Center for Quantum Sciences, Shanghai 201315, China
}%

\author{D. L. Feng}%
\email{dlfeng@fudan.edu.cn}

\affiliation{%
Laboratory of Advanced Materials, State Key Laboratory of Surface Physics, and Department of Physics, Fudan University, Shanghai 200438, China
}%
\affiliation{%
Shanghai Research Center for Quantum Sciences, Shanghai 201315, China
}%
\affiliation{%
Collaborative Innovation Center of Advanced Microstructures, Nanjing 210093
}%
\affiliation{%
Hefei National Laboratory for Physical Science at Microscale, CAS Center for Excellence in Quantum Information and
Quantum Physics, and Department of Physics, University of Science and Technology of China, Hefei 230026
}%

\date{\today}

\begin{abstract}
Filled-skutterudite heavy fermion (HF) compounds host rich ground states depending on the $f$ electron configurations. CeOs$_4$Sb$_{12}$ (COS) with Ce $4f^1$, and PrOs$_4$Sb$_{12}$ (POS) with Pr $4f^2$ configurations show distinct properties of Kondo insulating and HF superconductivity, respectivity. We unveiled the underlying microscopic origin by angle-resolved photoemission spectroscopy studies. Their eV-scale band structure matches well, representing the common characters of conduction electrons in $R$Os$_4$Sb$_{12}$ systems ($R$ = rare earth). However, $f$ electrons interact differently with conduction electrons in them. Strong hybridization between conduction electrons and $f$ electrons is observed in COS with band dependent hybridization gaps, and the development of Kondo insulating state is directly revealed. Although the ground state of POS is a singlet, finite but incoherent hybridization exists due to Kondo scattering with the thermally excited triplet crystalline electric field (CEF) state. Our results help to understand the intriguing properties in COS and POS, and provide a clean demonstration of the microscopic differences in HF systems with $4f^1$ and $4f^2$ configurations.


\end{abstract}

\pacs{Valid PACS appear here}
\maketitle


The interplay between localization and itineracy in heavy fermion (HF) systems gives rise to rich phase diagrams and physical properties highly sensitive to various tuning parameters\cite{CeIn3_phase,quantum_critical,RMP}.
It has been studied in Ce$T$In$_5$ ($T$~=~transition metal) that, by tuning the conduction electrons through varying $T$ among Co, Rh and Ir, minute changes in the hybridization between conduction electrons and $f$ electrons ($c$-$f$ hybridization) lead to distinct ground states\cite{CeCoIn5,CeRhIn5,CeIrIn5}.
Alternatively, tuning the $f$ electrons by replacing rare earth elements also induces distinct physical properties\cite{LaCoIn5,PrCoIn5,RCoIn5,ErCoIn5}, but the underlying physics is merely explored.
Moreover, unlike the intensively studied $4f^1$ system, experiments on the microscopic electronic behavior in $4f^2$ HF systems remain a missing piece.

Rare-earth compounds in filled-skutterudite structure ($RT_{4}X_{12}$, $R$~=~rare-earth compounds, $X$~=~pnictogen) offer an ideal playground in studying the Kondo physics with varied $f$ electrons, which host various ground states and novel physical properties, including superconductivity\cite{POS_HF_transport}, magnetism\cite{Ferro_EuTSb}, non-Fermi liquid behavior\cite{Non-Fermi_CePtGe,Non-Fermi_CeRuAs}, and semiconducting behavior\cite{COS_Kondo_Insulating,COS_Semiconducting}.
Particularly, unconventional superconductivity with transition temperature of 1.86~K and Kondo insulating behavior with T$_K$ around 90~K emerge out of the same lattice $R$Os$_4$Sb$_{12}$, in PrOs$_{4}$Sb$_{12}$ (POS)\cite{POS_transport_itinerant,POS_HF_transport} and CeOs$_{4}$Sb$_{12}$ (COS)\cite{COS_Kondo_Insulating}, respectively.
After almost two decades of studies, their key parameters in Kondo physics are controversial and the origin of the distinct ground states remain elusive. In POS,
HF superconductivity was evidenced by the large specific heat jump at superconductivity transition\cite{POS_transport_itinerant}, the nodal superconducting gap\cite{POS_Penetration_depth,POS_nodalSCSC,POS_nodal3} and the time reversal symmetry breaking\cite{POS_TRB}.
However, the cyclotron effective mass measured by de Haas-van Alphen is only 2.4$\sim$7.6 $m_{\rm{e}}$\cite{POS_dHvA}, questioning the participation of $f$ electrons in the ground states. Singlet ground state\cite{POS_Quadrupolar_exp,POS_Field} and non-monotonic temperature dependence of effective mass\cite{POS_NMR} were reported, which cannot be explained by the conventional magnetic Kondo effect. These exotic behaviors were suspected to involve extra degrees of freedom from the Pr 4$f^2$ configuration, like the quadrupolar fluctuations\cite{POS_Quadrupolar_Theory,POS_Quadrupolar_exp}, whose effect on the electronic structure has not been identified. Meanwhile, COS in 4$f^1$ configuration hosts a ground state of antiferromagnetic (AFM) semiconducting phase with an enhanced effective mass\cite{COS_Kondo_Insulating}. The reported Kondo gap ranges from ∼1~meV to 70~meV \cite{COS_Kondo_Insulating,COS_optical_conductivity,COS_neutron}, which needs to be clarified by further electronic structure measurements. Moreover, the theoretically proposed Fermi surface (FS) nesting\cite{COS_nesting,COS_nesting2} and nontrivial topological nature\cite{COS_Topo} in COS call for a systematic scrutiny from electronic structures perspective.

Angle-resolved photoemission spectroscopy (ARPES) has played a key role in studying the electronic structure in HF systems with layered structure\cite{CeCoIn5,CeIrIn5,CeRhIn5,Ce2PdIn8,CeCoGeSi}. However, the three dimensional structure of filled skutterudites[Fig.~1(a)] poses significant challenges to the surface sensitive ARPES studies.
In this study, combining the bulk sensitive soft X-ray (SX) ARPES and high resolution vacuum ultraviolet (VUV) ARPES studies, both with micrometer-sized beam spots, we successfully obtained the momentum dependent electronic structures of POS and COS, and directly revealed the distinctive interplay of localization and itineracy.

\begin{figure}[t]
    \centering
    \includegraphics[bb=0 0 250 250,width=8.6cm]{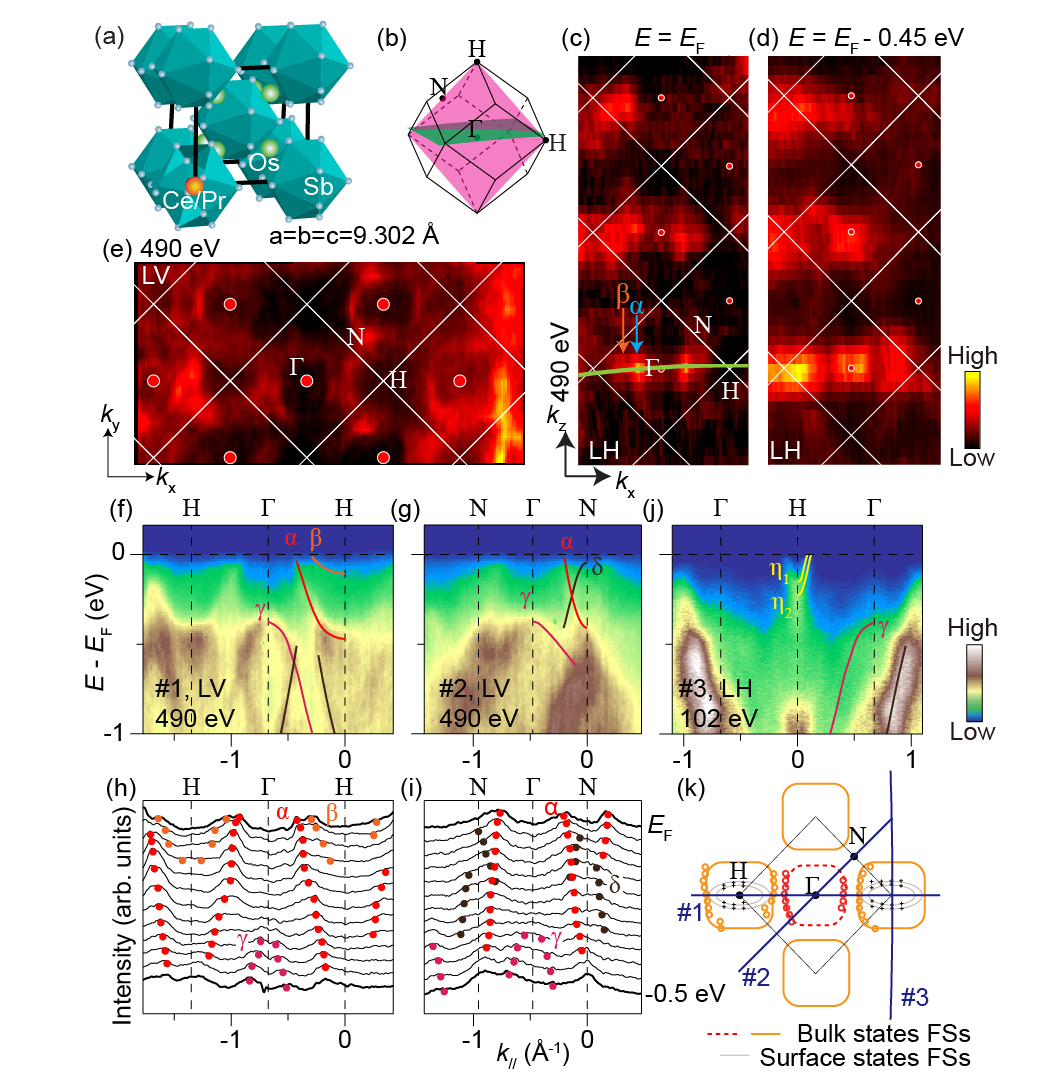}
    \caption{ (a) Crystal structure of the filled skutterudites. (b) Sketches of the Brillouin zone. The momentum spaces sampled in panels c (magenta) and e (green) are illustrated. (c) Photoemission intensity map  of POS in the $k_{\rm{x}}$-$k_{\rm{z}}$ plane measured with photons ranging from 400 to 800~eV, integrated over the energy window of $E_{\rm{F}}$~$\pm$~100~meV. (d) The same as (c) but integrated over [$E_{\rm{F}}$~-~0.55~eV, $E_{\rm{F}}$~-~0.35~eV]. (e) Photoemission intensity map  in the $k_{\rm{x}}$-$k_{\rm{y}}$ plane integrated over  $E_{\rm{F}}$~$\pm$~100~meV, collected with 490~eV photons (Green line in (c)). (f), (g) Photoemission spectra along H-$\Gamma$-H and N-$\Gamma$-N directions, respectively. (h)-(i) Momentum distribution curves (MDCs) correspond to (f) and (g), respectively. (j) Photoemission intensity along $\Gamma$-H-$\Gamma$ directions taken with 102~eV photons. (k) Sketches of FSs from both bulk states and surface states. Fermi crossings are determined by MDCs measured with 490~eV photons (open circles) and 102~eV photons (crosses). The momenta of
   the data shown in panels (f)-(j) are illustrated. All data were taken on POS at 19~K.}
\end{figure}

\begin{figure}[t]
    \centering
     
    \includegraphics[bb=0 0 300 200,width=8.6cm]{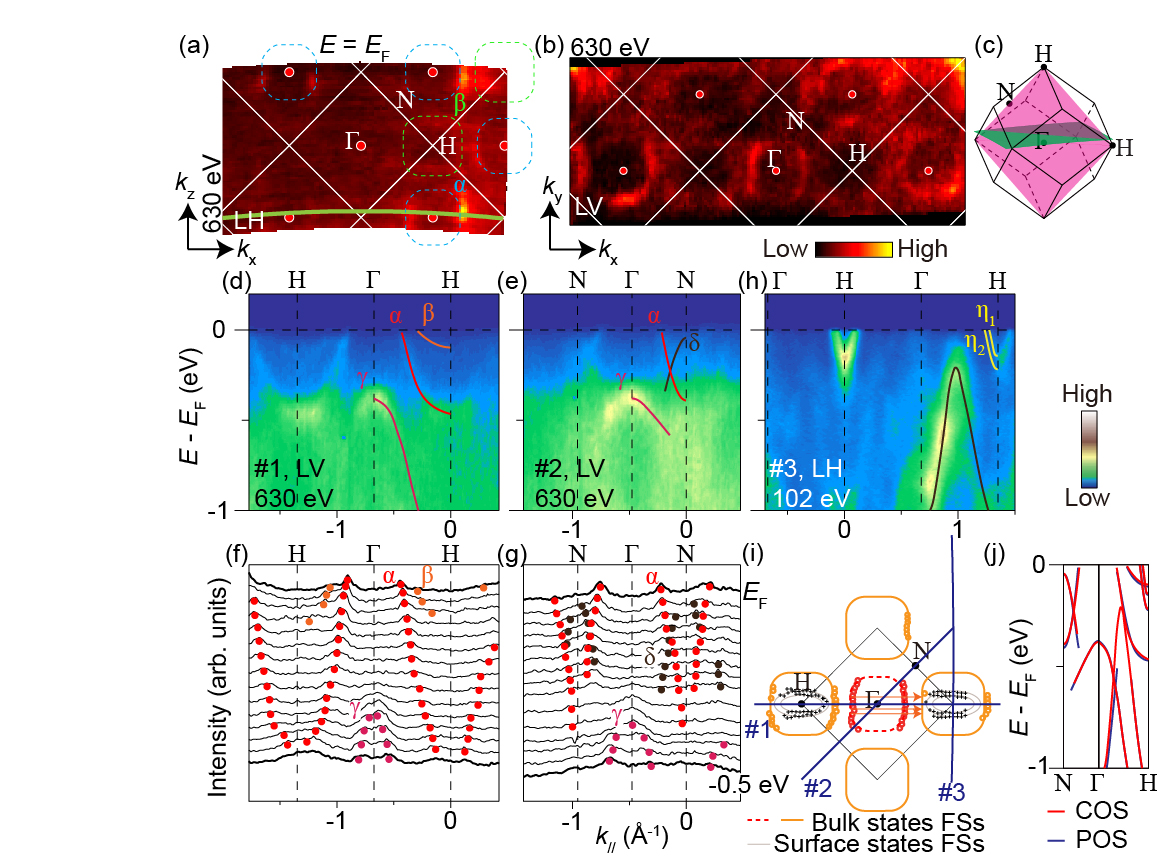}
    \caption{(a) Photoemission intensity map of COS in the $k_{\rm{x}}$-$k_{\rm{z}}$ measured with photons range from 615 to 780~eV. (b) Photoemission intensity map in the $k_{\rm{x}}$-$k_{\rm{y}}$ plane measured with 630~eV photons (Green line in (a)).  Both maps were integrated over the energy window of $E_{\rm{F}}$~$\pm$~100~meV. (c) Sketches of the Brillouin zone and the momentum spaces sampled in panels a (magenta) and c (green). (d)-(g) Photoemission spectra and the corresponding MDCs along H-$\Gamma$-H and N-$\Gamma$-N directions, respectively. Data were measured with 630~eV photons.  (h) Photoemission intensity along $\Gamma$-H direction taken with 102~eV photons. (i) Sketches of FSs from both bulk states and surface states. Fermi crossings are determined by MDCs taken with 630~eV photons (open circles) and 102~eV photons (crosses). Data in panels (a)-(i) were all taken at COS at 19~K. (j) Comparison of the band dispersions of POS and COS. }
\end{figure}

Single crystals of COS and POS were grown by powder metallurgical methods with high quality [Supplementary Fig.~S1\cite{supplementary}]. SX-ARPES data were taken at the beamline Advanced Resonant Spectroscopies of Swiss Light Source and VUV APRES studies were conducted at Diamond Light Source and Stanford Synchrotron Radiation Lightsource [Supplementary Methods\cite{supplementary}].

\begin{figure*}[t]
    \centering
    \includegraphics[bb=0 0 500 125,width=17cm]{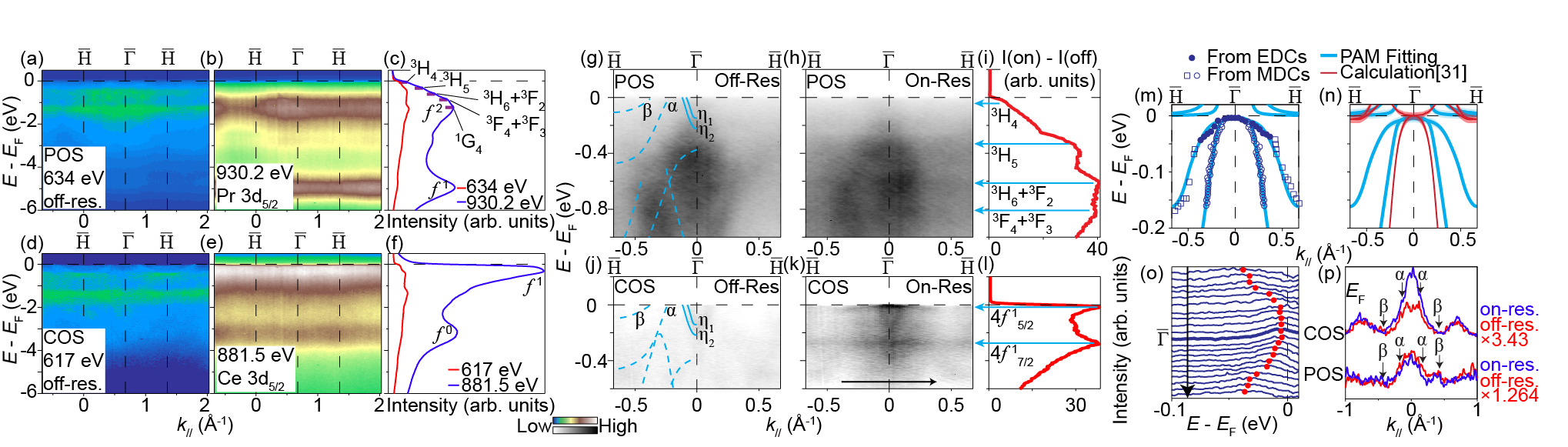}
    \caption{(a)-(b) Photoemission spectra of POS taken at 19~K with Pr M edge off-resonance (634~eV) (a) and on-resonance (930.2~eV) photons (b), which are equivalent in the k$_z$ periodicity. (c) Angle-integrated EDCs of (a) and (b). (d)-(f), Same as (a)-(c), but measured on COS with Ce M edge off-resonance (617~eV) and on-resonance (881.5~eV) photons. (g), (h) Photoemission intensity distributions of POS at 6.5~K taken with Pr N edge off-resonance (120.5 eV) and on-resonance (124.2 eV) photons, respectively. Dashed lines (not clear in this spectrum) and solid lines correspond to dispersions determined in Fig.~1.    (i) The intensity difference of the EDCs integrated over the momentum range in panels (g) and (h). (j)-(l) The same as (g)-(i) but for COS samples near N edge with off-resonance (118~eV) and on-resonance (121.9~eV) photons. (m) Results of PAM fittings to COS illustrating the $c$-$f$ hybridization. Blue circles and squares are obtained by fitting individual MDCs or EDCs. (n) A comparison between results of PAM fittings (blue) and calculation (red) in ref.~31. Blue and red shadows stand for the spectral weight from Ce 4$f$ electrons. (o) EDCs with the momenta shown by the arrow in panel (k). (p) MDCs at $E_{\rm{F}}$ of the data shown in panels (g)-(h) and (j)-(k).}
\end{figure*}

SX-ARPES measurements on POS using 400~eV to 800~eV photons reveal the FSs in the $k_{\rm{x}}-k_{\rm{z}}$ plane [Fig.~1(c)], which match the periodicity of the Brillouin zones [Figs.~1(c)-1(d)], confirming the bulk origin of the measured bands. Similar FSs are observed at the equivalent $k_{\rm{x}}-k_{\rm{y}}$ plane measured constant photon energy mapping [Fig.~1(e)], consistent with the body-centered cubic structure. Along $\Gamma$-H, band $\alpha$ and $\beta$ cross the $E_{\rm{F}}$, forming a hole pocket at $\Gamma$ and an electron pocket at H, respectively [Figs.~1(f) and~1(h)].
There is a hole-like band (noted as $\delta$) around the N point, with its band top barely touching $E_{\rm{F}}$ [Figs.~1(g) and~1(i)]. The photoemission spectra measured with VUV photons show distinct band structure from that taken with soft X-ray photons, demonstrating the existence of surface states.
Two electron bands $\eta_{1}$ and $\eta_{2}$ cross $E_{\rm{F}}$ near H [Fig.~1(j)], forming two elliptical pockets [Fig.~1(k)]. Their non-dispersive behavior along $k_{\rm{z}}$ [Supplementary Fig.~S2\cite{supplementary}] confirms their surface origin.

The 3D electronic structure of COS is generally identical to that of POS. As shown by the photoemission maps at $k_{\rm{x}}-k_{\rm{z}}$ plane and $k_{\rm{x}}-k_{\rm{y}}$ plane [Figs.~2(a)-2(b)] , the FSs consist of a hole pocket ($\alpha$) centered at $\Gamma$ and an electron pocket ($\beta$) centered at H.
The photoemission spectra along high symmetric directions of $\Gamma$ - H and $\Gamma$ - N [Figs.~2(d)-2(g)] highly resemble those of POS. Moreover, surface states $\eta_{1}$ and $\eta_{2}$ are also observed around H with 102~eV photons in COS [Fig.~2(h)], which shows two fold symmetry in $k_{\rm{x}}-k_{\rm{y}}$ plane and no dispersion along $k_{\rm{z}}$ [Supplementary Figs.~S4-S5\cite{supplementary}].
As summarized in Fig.~2(j), COS and POS show similar bandwidth, indicating the same high energy behavior in $R$Os$_4$Sb$_{12}$.

In order to detect the $f$-electron behavior contributed by Pr (Ce) in POS (COS), we conducted resonant ARPES measurements at both the M edges and N edges of Pr (Ce) elements.
The resonant photon energies are determined by X-ray absorption spectroscopy measurements [XAS, Supplementary Fig.~S6\cite{supplementary}]. In POS, resonant enhancement of photoemission signal is observed near $E_{\rm{F}}$ and at binding energies ($E_{\rm{B}}$) around 5~eV [Figs.~3(a)-3(b)],
which correspond to 4$f^{2}$ final states and 4$f^{1}$ final states, respectively [Fig.~3(c)]\cite{PrFeGe}. As the 4$f^{2}$ final state comes from the photoemission process where the $4f$ photon hole at 4$f^{1}$ final state is immediately recombined by a conduction electron\cite{HufnerTextBook}, the detectable intensity of 4$f^{2}$ final states corresponds to the hybridization from conduction electrons, indicating the finite $c$-$f$ hybridization in POS.
The photoemission resonance peak of the 4$f^2$ final states shows several sub-levels,  manifested as the flat bands in the Pr 4$d$ $\rightarrow$ 4$f$ resonant spectrum from VUV ARPES measurement with better energy resolution [Figs.~3(g)-3(i)]. The energies of the sublevels can be well accounted for by the multiplets of the 4$f^2$ configuration from theoretical calculations\cite{PrFeGe}. Nevertheless, ARPES spectra suggest the absence of well defined quasiparticle dispersion near $E_{\rm{F}}$ [Fig.~3(h)].

In COS, resonant enhancements of photoemission signals are observed at $E_{\rm{B}}$ = 3~eV and near $E_{\rm{F}}$, corresponding to the 4$f^0$ and 4$f^1$ final states respectively [Figs.~3(d)-3(f)]. Unlike the similar resonance magnitude of Pr 4$f^1$ and Pr 4$f^2$ final states in POS, the resonance of Ce 4$f^1$ near $E_{\rm{F}}$ is much stronger than the Ce 4$f^0$, indicating much stronger $c$-$f$ hybridization in COS.
At the Ce 4$d$ $\rightarrow$ 4$f$ resonant photon energy of 121.9~eV, resonant enhancement of two sharp flat bands at $E_{\rm{F}}$ and $E_{\rm{B}}$ = 0.25~eV are resolved, corresponding to the Ce 4$f^{1}_{5/2}$ and Ce 4$f^{1}_{7/2}$ final states, respectively [Figs.~3(j)-3(l)].
In contrast to the weak resonance behavior at $E_{\rm{F}}$ in POS, the Ce 4$f^{1}_{5/2}$ states in COS show a sharp coherence peak [Fig.~3(k)] and  heavy quasiparticle dispersion [Fig.~3(o)].
Intriguingly,  although the spectra from VUV photons are mainly contributed by the electron-like surface states $\eta_{1}$ and $\eta_{2}$, the hybridized bulk bands are still resolved
thanks to the resonant enhancement of the 4$f$ components. The heavy band shows a hole-like dispersion with band top at the Brillouin zone center [Fig.~3(o)].
At $E_{\rm{F}}$, the spectral weight of COS is enhanced at momenta inside the pocket $\alpha$ and outside the pocket $\beta$~[Fig.~3(p)] and the flat band dispersion smoothly connects to bulk bands $\alpha$ and $\beta$ from spectra taken with 630~eV and 200~eV photons [Fig.~3(m)].
By fitting the hybridized band to the mean field theory of the periodic Anderson model (PAM)\cite{PAM}, the energy dispersion can be given by
$E^{\pm}$=$\frac{\varepsilon_{f}+\varepsilon(k)\pm\sqrt{(\varepsilon_{f}-\varepsilon(k))^{2}+4\mid{V_{\rm{k}}}\mid^{2}}}{2}$
where $\varepsilon_{f}$ is the energy of the renormalized $f$-level and $\varepsilon(k)$ is the dispersion of the bare band of conduction electrons. At 6.5~K, we get $f$-level near $E_{\rm{F}}$ and $\mid{V_{\rm{k}}}\mid$ of 22$\pm{4}$~meV for $\alpha$ and 28$\pm{4}$~meV for $\beta$, corresponding to a direct gap of 44$\pm{8}$~meV for $\alpha$ and 56$\pm{8}$~meV for $\beta$, larger than that in CeCoIn$_5$\cite{CeCoIn5}, suggesting stronger $c$-$f$ hybridization in COS.

The development of heavy quasiparticle in COS is revealed by temperature dependent ARPES measurements at the Ce N edge. The spectral weight of $f$ states, which is not visible at 147~K, gradually increases from 147~K to 13~K [Figs.~4(a)-4(f)], indicating the increasing $c$-$f$ hybridization. At 13~K, a heavy band contributed by the $c$-$f$ hybridized electrons is well developed [Fig.~4(a)], with an emerging peak near $E_{\rm{F}}$ in the integrated EDCs [Fig.~4(g)]. As temperature decreases, the peak intensity increases monotonically, with no sharp transition [Fig.~4(i)]. Such a smooth crossover is typical in various HF systems\cite{CeIn3, Ce2PdIn8, CeCoGeSi, CeRhIn5, CeCoIn5, CeIrIn5} [Supplementary Fig.~S7\cite{supplementary}]. A gap opening due to $c$-$f$ hybridization could be observed [Fig.~4(h)], accounting for the developmemt of the Kondo insulating behavior in COS at low temperature.

\begin{figure}[t]
    \centering
    \includegraphics[bb=0 0 250 220,width=8.6cm]{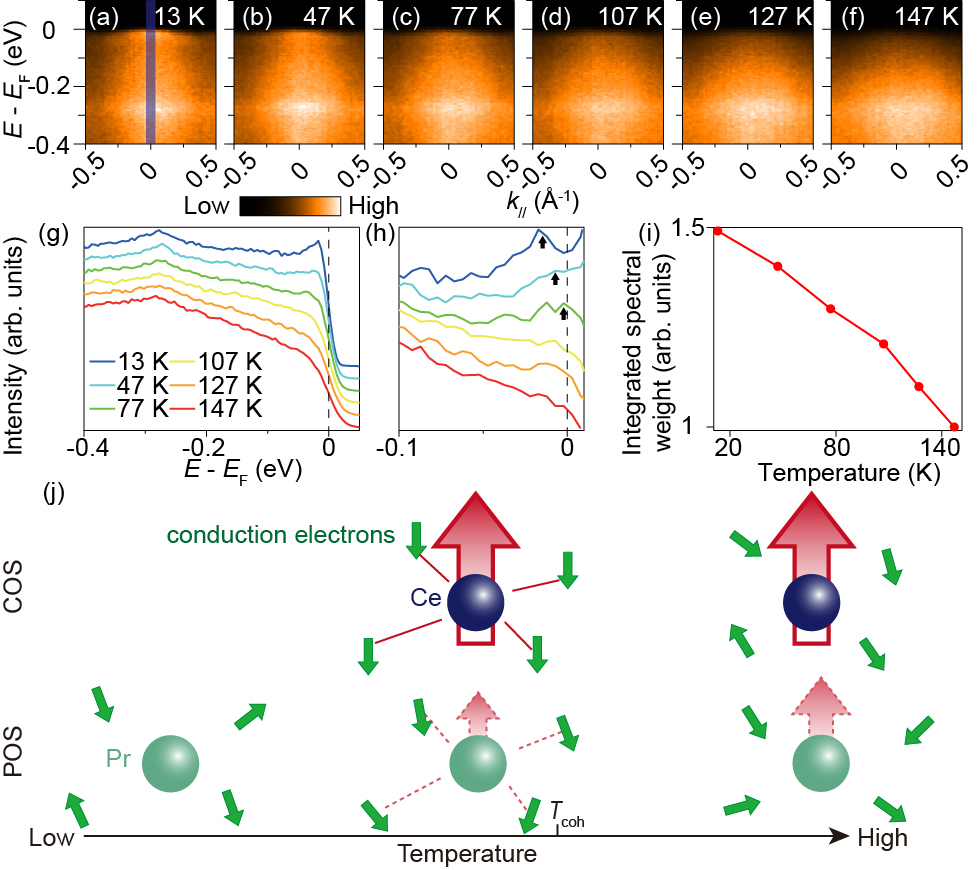}
    \caption{ (a)-(f) Resonant (121.6~eV, Ce N edge) photoemission spectra taken along $\bar{\Gamma}$-$\bar{\rm{H}}$ at different temperatures in COS. (g) EDCs integrated within the blue shadow area in (a) at different temperatures. (h) EDCs in (g) divided by the resolution convolved Fermi-Dirac function at corresponding temperatures as a typical way to compensate for the cutoff near $E_{\rm{F}}$ by Fermi-Dirac distribution. Positions of peaks are denoted by arrows. (i) Integrated spectral intensity over [$E_{\rm{F}}$ - 0.1, $E_{\rm{F}}$ + 0.1~eV], normalized by the intensity at 147~K. (j) Illustration of the magnetic Kondo screening in COS and POS evolving as a function of temperature. The green arrows indicate the spins of conduction electrons. The red arrows indicate the magnetic moments of Ce/Pr ions, with the size illustrating the average magnitude of the magnetic moments.}
\end{figure}

Our results reveal the distinctive characters of how the conduction electrons interact with $4f^1$ and $4f^2$ electrons. The observed electronic structure of POS and COS at eV-scale are strikingly similar, despite of the different rare-earth elements Pr and Ce. The difference between POS and COS is reflected by the $c$-$f$ hybridization at meV-scale near $E_{\rm{F}}$, as resolved in our resonant ARPES data. In COS, $c$-$f$ hybridization is observed and gets coherent at low temperature [Fig.~4(a)], which are hallmarks of the magnetic Kondo screening of conduction electrons to the magnetic moments of Ce 4$f^1$ orbital. In POS, the detection of 4$f^2$ final state in our experiments indicates the finite itineracy of Pr 4$f$ electrons and the presence of Kondo screening. However, the non-magnetic singlet ground state in POS alone\cite{POS_INS,POS_INS4,POS_INS5,POS_INS6} cannot host a magetic Kondo effect. Note that, as suggested by previous neutron scattering studies, just 0.7~meV above the $\Gamma_1$ ground state of Pr 4$f^2$ lies a triplet magnetic state $\Gamma_4^{(2)}$, forming a $\Gamma_1-\Gamma_4^{(2)}$ pseudo-quartet\cite{POS_review_2007,POS_INS2,POS_INS,POS_INS3,POS_INS5,POS_INS6,Lecture_Fazekas, CEF_cal}. Therefore, there would be a fair amount of Pr atoms thermally excited to the magnetic $\Gamma_4^{(2)}$ state at the measuring temperatures of ARPES studies (occupancy in $\Gamma_4^{(2)}$ state to that in $\Gamma_1$ state is 0.29 at 6.5~K and 0.66 at 19~K based on Boltzmann distribution), which leads to magnetic moments at the Pr ions. Given that these temperatures are below typical coherence temperature ($T_{\rm{coh}}$),  finite $c$-$f$ hybridization through exchange interaction between itinerant electrons and thermally exited moments is plausible, as indeed observed [Fig.~3(b)]. Considering that a large Van Vleck contribution is identified in magnetic susceptibility\cite{POS_transport_itinerant,POS_NMR}, there should be a considerable weight of $O_h$ triplet $\Gamma_4$\cite{Otsuki2005}, which favors antiferromagnetic in relative to ferromagnetic interaction\cite{Otsuki2005}. Therefore, the observed $c$-$f$ hybridization indicate finite Kondo interaction in POS due to the special CEF splitting.
Note that unlike the coherent Kondo state in COS, the Kondo effect in POS is incoherent with no well-defined dispersive quasiparticles [Fig.~3(h)]. This can be explained as the magnetic moments of Pr are transient due to thermal excitation, in contrast to the static Ce moments.
With decreasing temperature, despite of the otherwise enhanced Kondo coherence and the magnetism from the intermediate state of hybridization, the prevailing factor is the quickly decreasing and eventually vanishing of the magnetic moments due to less thermally-populated $\Gamma_4^{(2)}$ triplet states. The competition between these effects would lead to a nonmonotonic strength of $c$-$f$ hybridization as a function of temperature as illustrated in Fig.~4(j). Indeed, previous nuclear magnetic resonance study suggests a non-monotonic temperature dependence of effective mass, with a maximum around 3~K\cite{POS_NMR}. This competition also plausibly explain the reduced cyclotron effective mass at extremely low temperature of 30~mK\cite{POS_dHvA}.

Our results also provide a microscopic basis for understanding the exotic properties in COS.

1. The hybridization gap in COS was controversial and even off by one order of magnitude in different measurements\cite{COS_Kondo_Insulating,COS_optical_conductivity,COS_magnetic_gap,COS_neutron}, while the quantitative determination of Kondo gaps with momentum resolution here helps to understand the different gap sizes measured by previous experiments\cite{COS_Kondo_Insulating,COS_optical_conductivity,COS_neutron}.
Despite of the relatively large direct hybridization gap from PAM fitting [Fig.3(m)], the indirect gap is in the order of 1~meV and the top of the hybridized band is only $\sim8\pm$5~meV below Fermi energy at $\Gamma$ in the raw data [Fig.3(o)].
The small indirect gap could correspond to the small gap $\sim$1meV determined by transport measurements\cite{COS_Kondo_Insulating}, while the two direct gaps with band dependent sizes explain the two-gap features at the order of 30~meV and 50~meV in previous neutron and optical conductivity studies\cite{COS_optical_conductivity, COS_neutron}.
The fact of large direct gap and small indirect gap constitutes the intriguing dual characters of Kondo insulating behavior and HF behavior in COS at low temperature\cite{COS_Kondo_Insulating}.

2. Parallel FS sections are observed for the conduction electrons, which follows the nesting condition q=(1, 0, 0)[Fig.~2(b)] suggested as a driving force of the antiferromagnetic ground state below 0.9~K\cite{COS_nesting,COS_nesting2, COS_AFM}. Nevertheless, the nesting condition is no longer met at low temperature with the gapped electronic structure due to $c$-$f$ hybridization, which
question on whether the remnant FSs can play roles in driving the antiferromagnetism.

3.  A novel topological Kondo insulating state was proposed in COS by theoretical calculations suggesting $d$-$f$ band inversion near $E_F$\cite{COS_Topo}. However, topological surface states at $\Gamma$ within the Kondo gap is not detected within the current experimental resolution. The surface states located at H in both COS and POS are topologically trivial.
Moreover, although the calculation and experimental electronic structure both show strong 4$f$ component near $E_{\rm{F}}$, clear discrepancies can be observed[Fig.~3(n)].
In the calculation, the band bottom of the electron pocket at H is just at $E_{\rm{F}}$, while the corresponding band $\beta$ in our data disperses down to $E_{\rm{B}}$ = 0.16~eV and induce a flat band dispersing towards $\Gamma$ through hybridizing with the $4f$ band.
At $\Gamma$, the theory shows a band touching at $E_{\rm{F}}$ with zero gap , while our data show a finite Kondo gap of 8 $\pm$ 5~meV [Fig.~3(n)]. The measured band structure calls for a revisit on the theoretical calculations with corrected band structure and the scrutiny of the $d$-$f$ band inversion, which is a prerequisite of possible topological properties.

To summarize, we have experimentally unveiled the band dispersion in filled-skutterudites COS and POS with different $f$ electron configurations.
The band dispersion at eV-scale are strikingly similar in POS and COS, indicating that with the common electronic structure contributed by pure Os-Sb framework.
At the meV-scale energy range near $E_{\rm{F}}$, resonant ARPES data show drastic difference between POS and COS in the $c$-$f$ hybridization. Strong band dependent $c$-$f$ hybridization is observed in COS, directly demonstrating the development of Kondo insulating states. POS shows weak but visible $c$-$f$ hybridization, which is incoherent without well-defined dispersive quasiparticles, which is distinct from COS but relates with the CEF induced sublevels of Pr $4f^2$ configurations.
Besides clarifying several controversies in COS and POS from the electronic structure perspective, our results provide a clean demonstration on how Kondo physics behaves differently in $4f^1$ and $4f^2$ configurations, causing drastically different ground states.

We gratefully acknowledge the experimental support of Dr. Y. B. Huang, Dr. P. Dudin, Dr. D. H. Lu, Dr. V. N. Strocov and Dr. J. Denlinger. We thank the Diamond Light Source for access to beamline I05, the Shanghai Synchrotron Radiation Facility for access to beamline 09U, the Stanford Synchrotron Radiation Lightsource for access to beamline 5-2 and the Advanced Light Source for access to beamline 4.0.3.
This work is supported in part by Science Challenge Project (No. TZ2016004), the National Natural Science Foundation of China  (Grants No. 11888101, No. 11704073, No. 11704074, No. 11922403, No. 11790310, and No. 11774307), the National Key R$\&$D Program of the MOST of China (Grants No. 2016YFA0300200, No. 2017YFA0303104 and No. 2017YFA0303004), Anhui Initiative in Quantum Information Technologies and Shanghai Municipal Science and Technology Major Project (Grant No. 2019SHZDZX01).

\bibliographystyle{apsrev4-1}

\end{document}